\documentclass[aps, prd, 12pt,superscriptaddress,preprint,nofootinbib,showpacs,oneside]{revtex4}
\usepackage[latin1]{inputenc}
\usepackage[english]{babel}
\usepackage{graphicx,color, amsmath,amssymb}
\usepackage[caption=false]{subfig}
\usepackage[bookmarks]{hyperref}

\newcommand{\rmd}{\text{\rm{d}}}
\newcommand{\rmi}{\text{\rm{i}}}
\begin{document}
\title{Affine Quantization and the Initial Cosmological Singularity}
\author{Micha\"el Fanuel} \email{michael.fanuel@uclouvain.be}
\author{Simone Zonetti} \email{simone.zonetti@uclouvain.be}
\affiliation{Centre for Cosmology, Particle Physics and Phenomenology (CP3), Institut de Recherche en Math\'ematique et Physique, Universit\'e catholique de Louvain, Chemin du Cyclotron 2, bte L7.01.01, B-1348, Louvain-la-Neuve, Belgium}
\preprint{CP3-12-16}
\begin{abstract}
The Affine Coherent State Quantization procedure is applied to the case of a FRLW universe in the presence of a cosmological constant. The quantum corrections alter the dynamics of the system in the semiclassical regime, providing a potential barrier term which avoids all classical singularities, as already suggested in other models studied in the literature. Furthermore the quantum corrections are responsible for an accelerated cosmic expansion. This work intends to explore some of the implications of the recently proposed ``Enhanced Quantization'' procedure in a simplified model of cosmology.
\end{abstract}
\maketitle
\section{Introduction}
The initial singularity problem is a long standing problem in modern cosmology. It is often believed that the effects of quantum gravity should provide an answer to this question. Renowned candidate theories for quantum gravity are loop quantum gravity and superstring theories. Loop quantum cosmology and gauge-gravity duality are possible avenues of exploration (see for example \cite{Bojowald} and \cite{Craps}, respectively).
However other alternative or complementary approaches could be conceived. Among them, affine quantization has been recently put forward in order to quantize gravity \cite{KlauderAffine,KlauderAffineRecent,KlauderPrimer}, but has also been studied previously in \cite{Isham1,Isham2}, while this approach was used to study a strong coupling limit of gravity in \cite{Pilati1,Pilati2,Pilati3}. It is certainly interesting to examine the implications of this proposal. In this work, we apply the Affine Coherent State Quantization program to the dynamics of the scale factor in the FLRW (Friedman-Lema\^itre-Robertson-Walker) framework for cosmology, inclusive of a cosmological constant. We notice that the quantum corrections provide in an natural way a potential barrier term and we analyse the semiclassical behaviour using the ``Weak Correspondence Principle''.\\
In Section \ref{sec:classicalmodel} we introduce the classical model and calculate the classical equations of motion. Section \ref{sec:affinequant} introduces the Affine Coherent State Quantization scheme and discusses the derivation of the Extended Hamiltonian. The equations of motion are also calculated and the effect of the quantum corrections is briefly discussed. In Section \ref{sec:numerical} the numerical solutions for the classical and semiclassical cases are studied and compared, while in Section \ref{sec:conclusions} we present our conclusions. 
\section{The classical model}\label{sec:classicalmodel}
In an earlier work \cite{KlauderAslaksen}, a toy model for gravitation was studied from the affine perspective and it was argued that the singularities of the classical solutions were regularized because of the quantum effects. In a more recent article \cite{KlauderAffine} the semiclassical behaviour of the one-dimensional Hydrogen atom was analysed and it was shown a potential barrier emerges at the scale of the Bohr radius resolving the Coulomb singularity. In a similar way, we suggest here a simple model of a FRLW universe with a cosmological constant and discuss the consequences of the Affine Quantization on the classical singularities.
We shall consider the action:
\begin{equation}\label{FRLW_action}
S=\alpha\int  dt \ \frac{1}{2} N(t)a^{3}\left [-\frac{1}{N^{2}(t)}\left (\frac{\dot{a}}{a}\right )^{2}-\frac{\Lambda}{3}+\frac{k}{a^{2}}\right ],
\end{equation}
where $a(t)$ is the scale factor, $\Lambda$ is the cosmological constant, $k$ is the geometric factor and the scale $\alpha$ ensures that the action has the right dimensions. In the following we will set $\alpha = 1$ for simplicity. The explicit choice of a time coordinate $t$ is emphasized by the presence of the lapse function $N(t)$.\\
As it is well known, classical solutions to this model, because of the constraints produced by diffeomorphism invariance, depend on the value of the factor $k$ and the sign of the cosmological constant. In particular de Sitter solutions ($\Lambda > 0$) are available for all values of $k$, while Anti-de Sitter solutions ($\Lambda<0$) are only possible with $\kappa=-1$. A vanishing cosmological constant, on the other hand, does not allow a solution for $\kappa=1$.
\subsection{Hamiltonian Formulation}
In what follows we will relabel $a(t)=q(t)$. Given the Lagrangian density in \eqref{FRLW_action} the corresponding Hamiltonian, in the gauge $N(t)=1$, reads
\begin{equation}
H_{0}(p,q)=-\frac{p(t)^2}{2 q(t)}-\frac{1}{2} \kappa  q(t)+\frac{1}{6} \Lambda q(t)^3,
\end{equation}
where $p$ is the conjugate momentum of $q$. The equations of motion are easily calculated as:
\begin{eqnarray}
& \frac{p(t)}{q(t)}+q'(t)=0 \\
 &2 p'(t)=-\Lambda  q(t)^2+\kappa-\frac{p(t)^2}{q(t)^2}
\end{eqnarray}
The Hamiltonian is constrained to vanish as per effect of the diffeomorphism invariance. The symplectic structure is given by the Poisson bracket $\{q,p\}=1$. The configuration space variable $q$ is constrained to stay strictly positive: $q>0$. At this stage quantizing the phase space with canonical commutation relations $[Q,P]=i\hbar$ would lead to difficulties of interpretation if the spectrum of the self-adjoint operator $Q$ is the real line, i.e. including negative eigenvalues. Actually it is possible to define the operators $P$ and $Q$ so that $[Q,P]=i\hbar$ and $Q>0$, however in this instance the operator $P$ will only be hermitian (symmetric) but not self-adjoint, namely $P$ may not be made self-adjoint by any choice of boundary conditions. Hence the exponential $\exp \rmi q P/\hbar$ will then not be an unitary translation operator, as can be shown easily \cite{KlauderAslaksen}. Thus the canonical operators are not suitable, and a new set of kinematical self-adjoint operators are needed. We will see that quantizing another algebra of operators constitutes an interesting alternative.
\section{Affine Coherent State Quantization}\label{sec:affinequant}
\subsection{Construction of Affine Coherent States}
A long time ago Affine Coherent States have been claimed to be useful in order to quantize gravity in its ADM (Arnowitt-Deser-Misner) formulation \cite{OverviewKlauder, Isham1, Isham2}. These states rely on the quantization of the ``$ax+b$'' affine algebra rather than the Heisenberg algebra. The major advantage for their use in a quantization of gravity is that they appropriately implement the condition of positive definiteness of the spatial metric. In the problem at hand we have a similar condition on the ``scale factor'': $q>0$.
In order to define the affine coherent states we introduce the affine variables $(q,d)$ by defining $d=qp$, which reparametrize the phase space $(q,p)$.
The \emph{affine} coherent states
\begin{equation}
|p,q\rangle=e^{\rmi p Q/\hbar}e^{-\rmi \ln(q/\mu)D/\hbar}|\eta\rangle\label{AffineStates}
\end{equation}
form an overcomplete basis of the Hilbert space and $\mu$ is a scale with dimension of length. The fiducial vector $|\eta\rangle$  satisfies the polarization condition 
\begin{equation}\label{Polarization}
\left [\frac{Q}{\mu}-1+\rmi  \frac{D}{\beta\hbar}\right] |\eta\rangle=0 ,
\end{equation}
with $\beta$ a free dimensionless parameter.
In particular one has:
\begin{eqnarray}\label{average}
\langle\eta|Q|\eta\rangle=\mu ,\\
\langle\eta|D|\eta\rangle=0 .
\end{eqnarray}
It is worth to notice that the condition \eqref{Polarization} is built by analogy with the canonical coherent states construction and provides a differential equation for the wave function of the fiducial state. Because the state $|\eta\rangle$  satisfies $0< \langle\eta|Q^{-1}|\eta\rangle <\infty$, the associated coherent states \eqref{AffineStates} admit a resolution of identity:
\begin{equation}
\mathbb{I}=\int \! \! \frac{dpdq}{2\pi\hbar C} \ |p,q\rangle\langle p,q|  ,
\end{equation}
where $C=\mu \langle\eta|Q^{-1}|\eta\rangle$. Subsidiarily, it should be underlined that a canonical coherent state construction would not be meaningful here, because the momentum operator $P$ may not be made self-adjoint on the half line, as it is well known from the von Neumann theorem and the deficiency indices theory.
We may now proceed to the \emph{affine} quantization of the Hamiltonian formulation.
In terms of the affine variables $q$ and $d=qp$ the Hamiltonian takes the form
\begin{eqnarray}
H_{0}(p,q)=-\frac{d^2}{2 q^3}-\frac{1}{2} \kappa  q+\frac{1}{6} \Lambda q^3,\label{CH}
\end{eqnarray}
which is suitable to apply to correspondance principle $q\to Q$ and $d\to D$.
The classical affine commutation relations $\{q,d\}=q$ are quantized as $[Q,D]=i\hbar Q$. The operators $D$ and $Q$ are conveniently represented in $x$-space by 
\begin{eqnarray}
&D f(x)& = -\rmi\hbar(x\partial_{x}+1/2) f(x)=-\rmi\hbar \ x^{1/2} \partial_{x}(x^{1/2}f(x))\label{D_repr}\\
&Q f(x)& = x f(x),
\end{eqnarray}
so that the interpretation of the algebra in terms of dilatations is now completely intuitive. 
For the sake of the consistency of the coherent state definition \eqref{AffineStates}, the operator $D$ represented above should be self-adjoint, while there is no difficulty to define properly the operator $Q$ and its domain. In order to thoroughly specify $D$, we require that the boundary term, originating from
\begin{eqnarray}
\langle \phi|D\psi\rangle-\langle D^{\dagger} \phi|\psi\rangle=-\rmi\hbar \int_{0}^{+\infty} \rmd x \ \partial_{x}[\phi^{*}(x)\ x\ \psi(x)],\label{BoundaryTermD}
\end{eqnarray}
vanishes. Because the wave functions $\psi(x)\in \text{Dom }  D$ and $\phi(x)\in \text{Dom } D^{\dagger}$ have to be square integrable on the half line, they should both verify, in particular, the condition 
\begin{eqnarray}
\lim_{x\to 0}x^{1/2}\psi(x)=0=\lim_{x\to 0}x^{1/2}\phi(x),
\end{eqnarray}
which means that, if $|\psi(x)|$ diverges at zero, one can find $\epsilon>0$ so that $|\psi(x)|$ diverges slowlier than $x^{-1/2+\epsilon}$ close to zero. Thanks to \eqref{BoundaryTermD}, we can actually notice that the domains of $D$ and $D^{\dagger}$ indeed coincide.

The self-adjoint operators $Q$ and $D$ appropriately realize the algebra $[Q,D]=i\hbar Q$. The representation theory of such algebra guarantees the existence of a unitary irreducible representation with the spectrum $Q>0$ \cite{Aslaksen}.
 The fiducial state is then described by the wave function
\begin{equation}\label{fiducial_wf}
\langle x|\eta\rangle=N (x/\mu)^{\beta-1/2}\exp(-\beta x/\mu),
\end{equation}
with $N=(2^{-2\beta}\beta^{-2\beta}\Gamma[2\beta]\mu)^{-1/2}$. It is easy to interpret the role of $\mu$ from \eqref{fiducial_wf} in a comparison of \emph{affine} and \emph{canonical} coherent states: in the case of the latter a parameter $\lambda_q$ sets the width of the Gaussian fiducial vector, as in
\begin{equation}
\langle x| \Omega\rangle =\left(\frac{\pi\hbar}{\lambda_{0}}\right )^{-1/4}e^{-\lambda_{0}x^{2}/2\hbar},
\end{equation}
which satisfies: $[P/\lambda_{p}-\rmi Q/\lambda_{q}]|\Omega\rangle=0$, with $\lambda_{0}=\lambda_{p}/\lambda_{q}$.
For simplicity both scales are usually used to define a unit system so that $\lambda_{p}/ \lambda_{q} =1$.\\
Therefore the parameter $\mu$ of \emph{affine} coherent states can be interpreted as the analogue of $\lambda_q$, since it sets the width of the fiducial wave function and the average value of $Q$ in the affine coherent states. Besides, if we wish to extend further the comparison, $\beta\hbar$ is the analogue of $\lambda_{p}\lambda_{q}$ as we may guess from \eqref{D_repr}: $D=Q^{1/2}PQ^{1/2}$. The existence of $\beta$ can be understood as an artifact of the representation. Different values of $\beta$ lead to different representations of the same physical states.\\
However it is possible to see that there is a lower bound on the value of $\beta$: if we require the matrix element $\langle\eta|Q^{-1} D Q^{-1} D \dots|\eta\rangle$ (containing a number $n$ (resp. $n-1$) of $Q^{-1}$ (resp. $D$) operators) and $\langle\eta|Q^{-n} |\eta\rangle$ to be finite, we are forced to have $\beta > n/2$. We emphasize that this lower bound on the value of $\beta$ is dictated by mathematical consistency and not by physical arguments. Besides this constraint, no other requirement is set on $\beta$ at this stage, hence it will be considered as a free parameter. We will see that the specific value of $\beta$ is irrelevant in determining the qualitative cosmological behaviour in the semiclassical regime.

\subsection{Quantization and the semiclassical regime}
One proceeds to quantization of the classical Hamiltonian \eqref{CH} by defining the quantum Hamiltonian as
\begin{equation}
\mathcal{H}'(Q,D)=-\frac{1}{2}Q^{-1}DQ^{-1}DQ^{-1}-\frac{1}{2} k Q + \frac{1}{6}\Lambda Q^3.
\end{equation}
The choice of operator ordering taken here is the one consistent with the Coherent State Quantization ``rule'', also called ``anti-Wick quantization''. In order to have a self-adjoint operator, the conditions on the domain of $K=Q^{-1}DQ^{-1}DQ^{-1}$ have to be specified. We should require that the boundary term
\begin{eqnarray}
\langle \phi|K \psi\rangle-\langle K^{\dagger} \phi|\psi\rangle=\hbar^{2} \int_{0}^{+\infty} \rmd x \ \partial_{x}[\psi(x)\frac{1}{x}\partial_{x}\phi^{*}(x)-\phi^{*}(x)\frac{1}{x}\partial_{x}\psi(x)],\label{BoundaryTermK}
\end{eqnarray}
vanishes. Hence we can choose to ask that $\psi(x)\in \text{Dom } K$ verifies
\begin{eqnarray}
\lim_{x\to 0}x^{-1}\psi(x)=\lim_{x\to +\infty}x^{-1}\psi(x)=0.\label{SelfAdjointK}
\end{eqnarray}
The functions $\phi(x)\in \text{Dom } K^{\dagger}$, as may be seen from \eqref{BoundaryTermK}, have to satisfy the same conditions \eqref{SelfAdjointK}. As a result, the domains of $D$ and $D^{\dagger}$ coincide.
Let us point out that the affine coherent states belong to the domain of $K$, whenever $\beta>3/2$. Namely the wave function of an affine coherent state reads
\begin{eqnarray}
\langle x| p,q\rangle =N  (x/\mu)^{\beta-1/2}(\mu/q)^{\beta}e^{-\beta x/q}e^{i p x/\hbar},
\end{eqnarray}
which verifies \eqref{SelfAdjointK} when $\beta>3/2$. The Hilbert space of states and the domain of the relevant operators being thoroughly identified, we may now try to take advantage of the coherent states to understand the dynamics.

The quantum dynamics of the model may be described by a Coherent State Path Integral but in a first stage of this work we are interested in the classical limit of the system as viewed by a macroscopic observer. The notion of geometry being difficult to interpret in a purely quantum theory of gravity we are tempted to consider a semiclassical quantity that could emerge from the quantum theory and be interpreted in a geometrical context. The Extended Classical Hamiltonian provides such a description. It is associated to a Coherent State $|p,q\rangle$ as
\begin{equation}\label{ext_class_h}
h(p,q)=\langle p,q|\mathcal{H}'(Q,D)|p,q\rangle,
\end{equation}
and should take into account quantum corrections while describing a semiclassical behaviour. We follow here the ``Weak Correspondence Principle'' as advocated by Klauder \cite{WeakCP}. Intuitively we would like that classical and quantum mechanics coexist as they do in the physical world. The weak correspondence principle allows us to consider quantum effects in a classical description of the world where we know that $\hbar$ takes a non-vanishing finite value.
The fundamental reason why \eqref{ext_class_h} is believed to incorporate quantum corrections is that it originates from the variational principle implementing the Schr\"odinger equation
\begin{equation}
S_{Q}=\int \rmd t \ \langle \psi(t)|\rmi\hbar\partial_{t} -\mathcal{H}'(Q,D)|\psi(t)\rangle,
\end{equation}
but where the ``restricted'' quantum action is varied only on the set of (affine) coherent states $|p(t),q(t)\rangle$ rather than the full space of quantum states. Because of their semiclassical features, we believe that the coherent states may be the only ones accessible to a classical observer. Consequently, this restricted action principle
\begin{equation} 
\begin{split}
S_{Q(R)}&=\int \rmd t \ \langle p(t),q(t)|\rmi\hbar\partial_{t} -\mathcal{H}'(Q,D)|p(t),q(t)\rangle\\
&=\int\rmd t \ \big[-q(t)\dot{p}(t)-h(p,q)\big],
\end{split}\label{RestrictedAction}
\end{equation}
gives a motivation for considering the equations of motion of $h(p,q)$ as a meaningful semiclassical approximation of the dynamics of the quantum system \cite{KlauderPrimer}. Finally, we underline here the noticeable result that, starting from an \emph{affine} quantized theory, the restricted action leads to a canonical theory \eqref{RestrictedAction}. \\
Making use of
\begin{equation}
\langle p,q|\mathcal{H}'(Q,D)|p,q\rangle=\langle \eta|\mathcal{H}'(\frac{q}{\mu}Q,D+p\frac{q}{\mu}Q)|\eta\rangle,
\end{equation}
we obtain
\begin{equation} \begin{split}
h(p,q)=&-\frac{\mu ^3}{2q^3}\left\langle Q^{-1}D Q^{-1}D Q^{-1}\right\rangle  -\\&-\frac{\mu p^2}{2q}\left\langle Q^{-1}\right\rangle - \frac{1}{2}\frac{\kappa }{\mu } q\langle  Q\rangle +\frac{\Lambda }{6}\frac{q^3}{\mu ^3}\left\langle Q^3\right\rangle.
\end{split}
\end{equation}
The required matrix elements can be easily calculated using \eqref{fiducial_wf} and \eqref{D_repr}, and read
\begin{eqnarray}
&\left\langle Q^{-1}D Q^{-1}D Q^{-1}\right\rangle = \mu^{-3}\gamma &\qquad \mbox{with: }\beta>3/2,\\
&\left\langle Q^{-1}\right\rangle = \mu^{-1}Z &\qquad \mbox{with: }\beta>3/2,\\
&\left \langle  Q\right \rangle = \mu &\qquad \mbox{with: }\beta>3/2,\\
&\left\langle Q^3\right\rangle = \mu^{3} \delta  &\qquad \mbox{with: }\beta>3/2,
\end{eqnarray}
where
\begin{eqnarray}\label{Zgammadelta}
\gamma=\frac{2 \beta ^3 \hbar ^2}{(3+4 (-2+\beta ) \beta )}>0,\label{gamma}\\
Z = \frac{2\beta}{2 \beta -1}>1,\\
\epsilon = \frac{4 \rmi \beta ^3 \hbar  \Gamma (2 \beta -3)}{\Gamma (2 \beta )},\\
\delta  = \frac{(1+\beta ) (1+2 \beta )}{2 \beta ^2}>1.\label{delta}
\end{eqnarray}
Note that both quantities are independent of the scale $\mu$ and finiteness of these matrix elements requires a lower bound on the value of $\beta$\footnote{The condition $\beta>3/2$ is already required by \eqref{SelfAdjointK} so that the coherent states belong to the domain of the quantum Hamiltonian.}.
Notwithstanding we have to emphazise that, once $\beta$ is chosen so that all matrix elements are finite, the value of $\gamma=||Q^{-1/2}DQ^{-1}|\eta\rangle||^{2}$ may never be negative. We stress that the value of the matrix elements given above should only be evaluated for $\beta>3/2$, while other values of $\beta$ lead to inconsistent results. To summarize, we find that the Extended Hamiltonian takes the form
\begin{eqnarray}
h(p,q)=-\frac{Z p(t)^2}{2 q(t)}-\frac{\gamma }{2 q(t)^3}+\frac{1}{6} \delta  \Lambda  q(t)^3-\frac{1}{2} \kappa  q(t). \label{QuantumCorrectedH}
\end{eqnarray}
Once again diffeomorphism invariance will require $h(p,q)=0$ to be enforced by the dynamics. It is remarkable that the dependency on the scale $\mu$ has been completely simplified. The classical limit \eqref{CH} of the Extended Hamiltonian is readily reproduced by taking simultaneously $\hbar \to 0$ and $\beta \to \infty$, while their product is kept constant $\hbar \beta \to \tilde{\beta}$. In this way we obtain that $Z\to1$ and $\delta \to 1$, while $\gamma\to0$.
\subsection{Qualitative analysis of the dynamics}
Interestingly the quantum corrections generate one unique new dynamical term in the Hamiltonian, proportional to $q^{-3}$. This contribution will naturally affect the dynamics for small values of the scale factor $q$. We can infer its behaviour by looking at the equations of motion for \eqref{QuantumCorrectedH}, which read:
\begin{eqnarray}\label{q_EOM}
&Z \frac{p(t)}{q(t)}+\dot{q}(t)=0,\\
&Z \frac{p(t)^2}{q(t)^2}+ \gamma\frac{3}{q(t)^4}+\delta  \Lambda  q(t)^2-\kappa +2 \dot{p}(t)=0.
\end{eqnarray}
As it is known in General Relativity the large scale gravitational dynamics,\emph{i.e.} $q \gg 0$, is dominated by the cosmological constant term: $\Lambda>0$ generates a repulsive force and determines an accelerated expansion while $\Lambda<0$ is responsible for an attractive force that, for example, can slow down cosmic expansion.\\
In the same way, as we can see from \eqref{q_EOM}, the small scale dynamics, \emph{i.e.} $q \ll 1$, will be dominated by the second term, proportional to $\gamma$. This quantity is always positive for $\beta>3/2$, hence it behaves as a small scale equivalent of a positive cosmological constant, generating a repulsive force when the universe contracts. In particular, as we will see by solving numerically the equations of motion, this quantum correction is able to keep the scale factor from vanishing, avoiding to reach big crunch singularities.\\
Furthermore the large scale behaviour is also modified: the constant $\delta$, defined in \eqref{delta}, multiplies $\Lambda$ and it is strictly greater than 1 for finite $\beta$, so that the effects of the cosmological constant are amplified for finite $\beta$ and the effective cosmological constant is $\delta \Lambda$. Finally we can see also that $Z>\delta>1$ for all $\beta$. Therefore even if different $\beta$'s label distinct quantum theories, the qualitative effects, as the avoidance of the classical singularity and the increased expansion rate, are universal.
\section{Numerical solution of the equations of motion}\label{sec:numerical}
To illustrate the claims of the previous section it is possible to consider numerical solutions to the classical and semiclassical equations of motion, comparing the behaviour of the scale factor with the same (or close enough) set of initial conditions for both regimes. This however is non trivial due to the presence of the diffeomorphisms constraints: to be able to have a meaningful comparison both \eqref{FRLW_action} and \eqref{QuantumCorrectedH} have to vanish simultaneously at any given time $t$.\\ Ideally we would like to solve the system of equations $H\left (p_0,q_0\right )=h\left (p_0,q_0\right )=0$ to (possibly) determine a unique set of initial conditions $(p_0,q_0)$ as functions of the parameters $\Lambda,\kappa,\beta$: this turns out to be possible only for the case of a de Sitter universe ($\Lambda>0$) with $\kappa = 1$. In a more pragmatic approach we will apply the following procedure in all possible combinations of $\Lambda \lesseqgtr 0$ and $\kappa=\pm1, 0$:
\begin{enumerate}
\item The parameters $\Lambda, \kappa$ and the initial value $q_0$ are fixed, identical for the classical and semiclassical cases, arbitrarily but allowing a solution of the constraints. The value of $\hbar$ is fixed to $\hbar=0.1$.
\item The initial value $p_0^c$ for the classical momentum is obtained from the constraint equation $H(p_0^c,q_0)=0$. 
\item The initial value $p_0^a$ for the momentum of the semiclassical (affine)regime is expressed as a function of $\beta$ by solving the constraint equation $h(p_0^a, q_0)=0$.
\item An optimal value of $\beta$r is determined by minimizing the difference $p_0^a(\beta)-p_0^c$. This has the purpose of providing initial conditions that are as close as possible for the two regimes.
\item The classical and semiclassical solutions are calculated numerically using the initial values just determined.
\item The quality of the numerical solutions is checked by requiring the classical and extended hamiltonians to have a numerical value smaller than $10^{-5}$ at all times.
\end{enumerate}
We can now look at the effects of the quantum corrections in all possible cases.
\begin{figure}
\subfloat[]{\label{fig:AdS_q} \includegraphics[width=.45\textwidth]{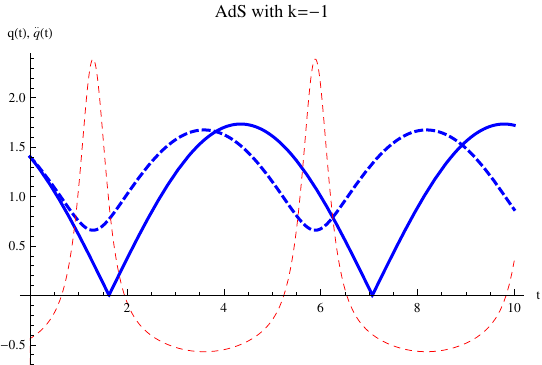}}
\subfloat[]{\label{fig:AdS_ps} \includegraphics[width=.45\textwidth]{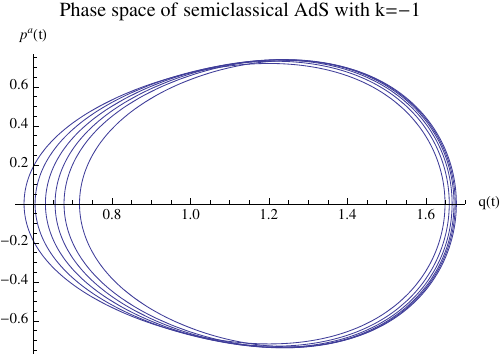}}
\caption{\label{fig:AdS} On the left: numerical solutions for an Anti-de Sitter universe with $\Lambda=-1$. On the left are plotted solutions for $q_0=1$. The continuous blue line is the plot of the classical solution for $q(t)$ (for the sake of comparison this case is plotted also after the first singularity is reached at $t\sim1$). The dashed lines refer to solutions in the semiclassical regime: $q(t)$ is plotted in blue while $\ddot{q}(t)$ is plotted in red. On the right: phase space trajectories for different initial conditions and different values of $\Lambda$.}
\end{figure}
\begin{itemize}
\item \textit{Closed ($\kappa=-1$) Anti-de Sitter universe} (Figure \ref{fig:AdS})\\
The specific form of the classical Hamiltonian, as mentioned earlier, allows the Hamiltonian constraint to be enforced only in the case $\kappa=-1$, in which the classical scale factor has a sinusoidal behaviour and it reaches $q=0$. With the inclusion of quantum corrections the singularity is avoided and the scale factor enters an infinite cycle of contractions and expansions, by effect of the $\gamma$ term. The frequency of these oscillations is increased with respect to the frequency of classical sinusoidal solution, due to the multiplicative constant $\delta>1$ which amplifies the effect of the cosmological constant.\\
In Figure \ref{fig:AdS_q} are plotted in blue the classical solution (continuous line), as a reference for the frequency, and the semiclassical solution (dashed line). Note how the minima of the semiclassical scale factor appear at earlier and earlier times with respect to the singularities of the classical $q$ for effect of the modified dynamics. From the plot of $\ddot{q}(t)$ (red dashed line in Fig. \ref{fig:AdS_q}) it is possible to visualize the important contribution to the cosmic acceleration provided by the $\gamma$ term.\\
The phase space trajectory for different initial values $q_0$, and therefore different values of $\beta$ and $p_0^a$, is plotted in Figure \ref{fig:AdS_ps} as an example of the independence of the behaviour from the specific value of $\beta$. With initial values for $q$ ranging from $q_0=0.8$ to $q(0)=1.3$ the required values for $\beta$ range from $\beta \sim 13$ to $\beta \sim 40$. In all cases the orbits are closed and the universe is bouncing.
\end{itemize}
\begin{figure}
\centering
\subfloat[]{\label{fig:dS_k=1} \includegraphics[width=.45\textwidth]{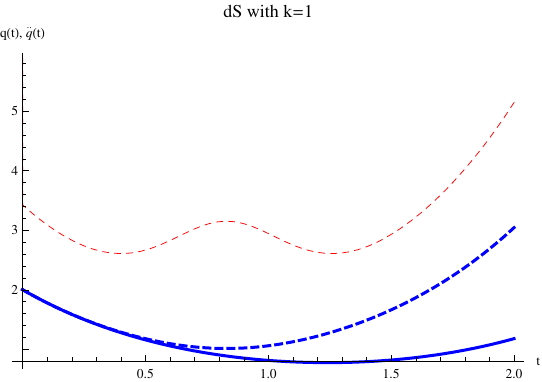}}
\subfloat[]{\label{fig:dS_k=0} \includegraphics[width=.45\textwidth]{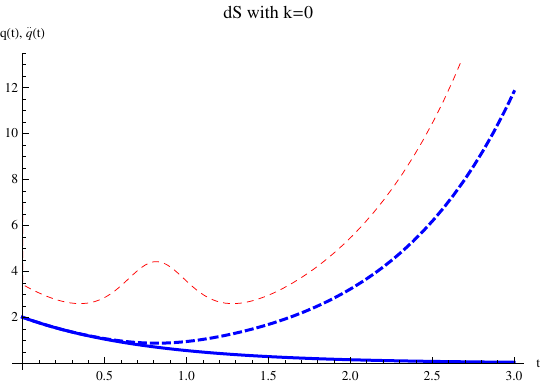}}\\
\subfloat[]{\label{fig:dS_k=-1} \includegraphics[width=.45\textwidth]{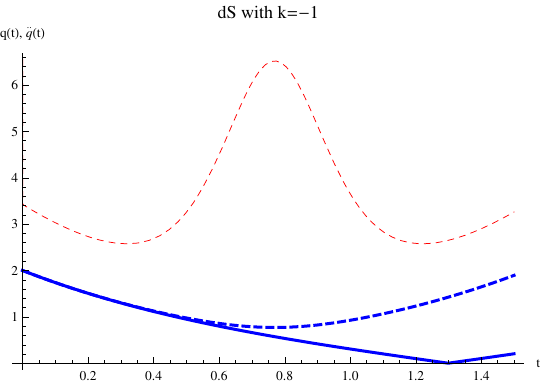}}
\subfloat[]{\label{fig:flat_k=-1} \includegraphics[width=.45\textwidth]{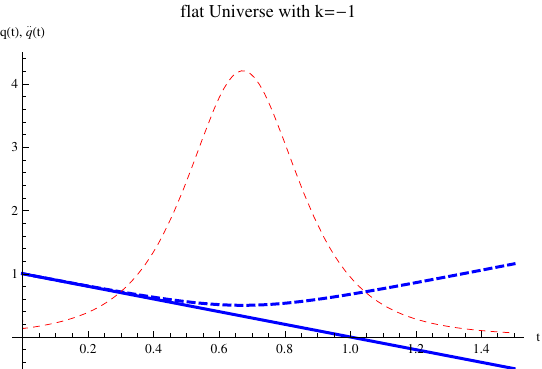}}
\caption{\label{fig:dS} Comparison between classical and semiclassical behaviour in the case of a de Sitter universe (with $\Lambda = 5,\ q_0=2$ for different values of the parameter $\kappa$) and a flat universe (with $\kappa=-1$). Continuous blue lines are classical solutions while dashed lines are semiclassical ones. Once more the dashed red line is $\ddot{q}(t)$ in the semiclassical regime.}
\end{figure}
\begin{itemize}
\item \textit{Open ($\kappa=-1$) de Sitter universe} (Figure \ref{fig:dS_k=1})\\
No classical singularity is present and the scale factor grows indefinitely, determining an accelerated expansion of the universe. The quantum correction however affects the dynamics speeding up the expansion and increasing the acceleration. Figure \ref{fig:dS_k=1} shows the plot for $q(t)$ and $\ddot{q}(t)$. Note again how the behaviour close around the minimum of $q(t)$ sees a substancial contribution from the $\gamma$ term.
\item \textit{$\kappa=0$ de Sitter universe} (Figure \ref{fig:dS_k=0})\\
At the classical level the scale factor decreases rapidly in a first phase and then slowly approaches $q=0$ asymptotically. In the semiclassical case the quantum correction is dominant after the initial rapid contraction and determines an highly accelerated expansion which avoids the classical singularity.
\item \textit{Closed ($\kappa=1$) de Sitter universe} (Figure \ref{fig:dS_k=-1})\\
The classical behaviour is singular, with a scale factor that reaches the singularity at finite times. The quantum correction once more avoids reaching $q=0$ and determines an accelerated expansion.
\end{itemize}
\begin{itemize}
\item \textit{$\kappa=0$ $\Lambda=0$ universe}
This is the only static solution for the classical model. To satisfy the classical constraint $p(t)$ has to identically vanish and $q(t)$ is in fact constant. However the extended hamiltonian $h$ is non-vanishing for any real set of allowed initial conditions, due to the presence of the positive constant $\gamma$:
\begin{eqnarray}
h(p,q)=-\frac{Z p(t)^2}{2 q(t)}-\frac{\gamma }{2 q(t)^3} \neq 0.
\end{eqnarray}
\item \textit{$\kappa=-1$ $\Lambda=0$ universe} (Figure \ref{fig:flat_k=-1})\\
 Again the classical behaviour is singular and is determined by the negative geometric factor $\kappa$, resulting in a linear, \emph{i.e.} constant velocity, approach of $q=0$. Also in this case quantum corrections are responsible for avoiding the singularity and inducing an expansion
\end{itemize}
Independence of these behaviours from the specific value of the parameter $\beta$ can be and has been tested successfully by repeating the analysis for different initial conditions.
\section{Conclusions}\label{sec:conclusions}
In this work we have studied the possibility of applying the Affine Coherent States Quantization scheme to a model of FRLW cosmology, in the presence of a cosmological constant. We considered a semiclassical regime, motivated by the ``Weak Correspondence Principle'' formulated by Klauder \cite{WeakCP}.\\
We found that the additional terms and multiplicative constants arising from the quantization of the dilation algebra profoundly changes the dynamics, independently from the specific value of the parameter $\beta>3/2$, which labels different quantum theories: the large scale dynamics is modified by an increased absolute value of the effective cosmological constant and the small scale dynamics is affected by a potential barrier generated by quantum corrections.
In the case of an open de Sitter universe, which already at the classical level exhibits no singularity and expands eternally, expansion is accelerated by a combination of small and large scale effects. The possibility of a connection with Dark Energy is worth investigating.
More interestingly all cases that possess a classical singularity, \textit{i.e.} $q\to0$, exhibit a non singular behaviour in the semiclassical regime and enter an expansion phase after reaching a minimal length at which the quantum dynamics is dominant. In the case of a closed Anti-de Sitter universe, in addition, the scale factor enters an infinite cycle of expansions and contractions.\\
These results, although limited to semiclassical considerations, provide additional support to the proposal of applying the affine quantization procedure in the approach of quantum gravity and quantum cosmology.  Further investigations should be put forward to fully understand the role of affine coherent states and the potential of this approach: for instance it would be interesting to see whether alternative choices for the fiducial vector are available and provide a similar behaviour; alternative ordering prescriptions can also be employed and their consistency has to be checked.

\acknowledgments
John Klauder is warmly thanked for his accurate comments and constant encouragements. It is a pleasure to acknowledge Jan Govaerts for his stimulating remarks and Christophe Ringeval for helpful discussions. The work of MF is supported by the National Fund for Scientific Research (F.R.S.-FNRS, Belgium) through a ``Aspirant'' Research fellowship and SZ benefits from a PhD research grant of the Institut Interuniversitaire des Sciences Nuclaires (IISN, Belgium). This work is supported by the Belgian Federal Office for Scientific, Technical and Cultural Affairs through the Interuniversity Attraction Pole P6/11.

\bibliographystyle{hunsrt}
\bibliography{references}
\end{document}